\documentstyle[12pt]{article}

\begin{document}

\title{{\bf Interacting open p-branes}}
\author{Sumio Ishikawa\thanks{E-mail: liszt@gradsun.ap.kagu.sut.ac.jp},
	Yasuhiro Iwama, \\
	Tadashi Miyazaki
	and
	Motowo Yamanobe\thanks{E-mail: yamanobe@gradsun.ap.kagu.sut.ac.jp}\\
	{\it Department of Physics,
	Science University of Tokyo} \\
	{\it Kagurazaka,
	Shinjuku-ku,
	Tokyo
	162,
	Japan}}
\date{December 1994}

\maketitle

\vspace{20mm}

\begin{flushleft}
\begin{small}
\end{small}
\end{flushleft}

\vspace{20mm}
\begin{abstract}
The Kalb-Ramond action,
derived for interacting strings through an action-at-a-distance force,
is generalized
to the case of interacting $p$-dimensional objects (p-branes)
in $D$-dimensional space-time.
The {\it open} p-brane version of the theory
is especially taken up.
On account of the existence of their boundary surface,
the fields mediating interactions between open p-branes are
obtained as {\it massive} gauge fields,
quite in contrast to massless gauge ones for {\it closed} p-branes.
\end{abstract}

\setlength{\baselineskip}{8.0mm}
\newpage

Some forty-five years ago,
Feynman and Wheeler\cite{feynman} presented an elegant idea
that the action-at-a-distance (AD) force gives us a gauge field (photon)
mediating electromagnetic interaction of a point particle.
Use was made of this idea by Kalb and Ranomd\cite{kalb}
to describe interacting strings\cite{kalb}\cite{ramond}.
They had the results that
the fields mediating the mutual interactions of strings are
either massless vector fields (for {\it closed} strings)
or massive vector ones (for {\it open} strings).

The present authors generalized Kalb-Ramond's method to the case
where extended objects of more than one-dimension are interacting
with one another\cite{yama}.
There {\it closed} p-branes were taken up and it was concluded that
it is the {\it massless} gauge field
that plays an essential role in the interaction of closed p-branes.

In the present paper,
the authors will confront with {\it open} p-branes
and construct the theory for the interacting open p-branes
{\it \`a la} Kalb-Ramond's AD view.
The existence of boundary surface in this case enforces us
to take the equations on surface {\it ad hoc} into account.
The difference here arises,
so that one is to have a {\it massive} gauge field,
as will soon be clear.

\vspace*{8.0mm}

Now let us start with the following action for two p-branes,
which is interacting each other through AD forces\cite{yama}
\begin{equation}
	S
	= S^{\rm free}_{a}
	+ S^{\rm free}_{b}
	+ \int d^{p+1}\xi_{a}\int d^{p+1}\xi_{b}
	R_{ab}(X_a,X_b,X_{a,i},X_{b,i}).                          \label{def:S}
\end{equation}
The p-brane is represented by the variables
$X^{\mu}(\xi)$ $(\mu=0,1,\ldots,D-1)$
in the flat $D$-dimensional space-time.
The world sheet of the p-brane is described by the variables
$\xi^{i}$ $(0\leq\xi^{i}\leq l^{i}$ $;$ $i=0,1,\ldots,p)$.
The subscript $a$ or $b$ labels each p-brane.
$S^{\rm free}$ is the action for a free p-brane,
and is taken to be the $(p+1)$-dimensional world volume
traced out by the p-brane
\begin{eqnarray}
	S^{\rm free}
	& = & -\int (-d\sigma \cdot d\sigma )^{\frac{1}{2}} \nonumber \\
	& = & -\int d^{p+1}\xi
	      (-\sigma\cdot\sigma)^{\frac{1}{2}}.             \label{def:Sfree}
\end{eqnarray}
The notations used in this paper are as follows:
\begin{eqnarray*}
	d\sigma \cdot d\sigma
	& \equiv &
	d\sigma^{\mu_{0}\mu_{1}\ldots\mu_{p}}
	d\sigma_{\mu_{0}\mu_{1}\ldots\mu_{p}}, \\
	\sigma\cdot\sigma
	& \equiv &
	\sigma^{\mu_{0}\mu_{1}\ldots\mu_{p}}
	\sigma_{\mu_{0}\mu_{1}\ldots\mu_{p}}, \\
	d\sigma^{\mu_{0}\mu_{1}\ldots\mu_{p}}
	& \equiv &
	d^{p+1}\xi \sigma^{\mu_{0}\mu_{1}\ldots\mu_{p}}, \\
	\sigma^{\mu_{0}\mu_{1}\ldots\mu_{p}}
	& \equiv &
	\frac{\partial(X^{\mu_0},X^{\mu_1},\ldots,X^{\mu_p})}%
	{\partial(\xi^0,\xi^1\ldots,\xi^p)}, \\
	d^{p+1}\xi
	& \equiv &
	d\xi^{0}d\xi^{1}\ldots d\xi^{p}, \\
	X_{,i}^{\mu}
	& \equiv &
	\frac{\partial X^{\mu}}{\partial\xi^{i}}.
\end{eqnarray*}
The $R_{ab}$ satisfy
\begin{equation}
	R_{ab} = R_{ba}                                           \label{Sym:R}
\end{equation}
for symmetry requirements of the action.
The variational principle for the variation
$X_{a}^{\mu}(\xi) \rightarrow X_{a}^{\mu}(\xi)+\delta X_{a}^{\mu}(\xi)$
immediately yields the equations of motion
\begin{equation}
	(p+1)D_a^{\mu_{1}\mu_{2}\ldots\mu_{p}}
	\left[
	\frac{\sigma_{a\mu_{0}\mu_{1}\ldots\mu_{p}}}%
	{(-\sigma_{a}\cdot\sigma_{a})^{\frac{1}{2}}}
	\right]
	=
	\int d^{p+1}\xi_{b}
	\left[
	(\frac{\partial}{\partial X_{a}^{\mu_0}}
	- \sum_{i=0}^{p}
	\frac{\partial}{\partial\xi_{a}^{i}}
	\frac{\partial}{\partial X_{a,i}^{\mu_0}})R_{ab}
	\right].                                                \label{eq:mo-1}
\end{equation}
Moreover,
for our {\it open} p-brane,
the boundary conditions on the surface add the following equations
\begin{equation}
	(p+1)
	\frac{\sigma_{a\mu_{0}\mu_{1}\ldots\mu_{p}}}%
	{(-\sigma_{a}\cdot\sigma_{a})^{\frac{1}{2}}}
	K_{ai}^{\mu_{1}\mu_{2}\ldots\mu_{p}}
	=
	- \int d^{p+1}\xi_{b}
	\frac{\partial R_{ab}}{\partial X_{a,i}^{\mu_0}},      \label{eq:sur-1}
\end{equation}
\begin{displaymath}
	(\xi_a^i = 0,l_a^i\hspace{1ex};\hspace{1ex}i=1,2,\ldots,p).
\end{displaymath}
Equation(\ref{eq:sur-1}) constitutes
one of the main characteristics of the open p-brane,
and it will play an essential role later.
Here and in what follows,
we define $K_{i}$ and $D$ by
\begin{eqnarray}
	K^{\mu_{1}\mu_{2}\ldots\mu_{p}}_{i}
	& \equiv &
	\frac{\partial\sigma^{\alpha\mu_{1}\mu_{2}\ldots\mu_{p}}}%
	{\partial X_{,i}^{\alpha}},                               \label{def:K}
	\\
	D^{\mu_{1}\mu_{2}\ldots\mu_{p}}
	& \equiv &
	\sum_{i=0}^{p}
	K^{\mu_{1}\mu_{2}\ldots\mu_{p}}_i
	\frac{\partial}{\partial\xi^{i}}.                         \label{def:D}
\end{eqnarray}
In the definition of $K_{i}$,
we do not sum over the index $\alpha$.
We assume that $R_{ab}$ depends on $X_{a,i}^{\mu}$ only through $\sigma_{a}$.
Under this restriction,
we rewrite Eq.(\ref{eq:mo-1}) as
\begin{equation}
	(p+1)D_a^{\mu_{1}\mu_{2}\ldots\mu_{p}}
	\left[
	\frac{\sigma_{a\mu_{0}\mu_{1}\ldots\mu_{p}}}%
	{(-\sigma_{a}\cdot\sigma_{a})^{\frac{1}{2}}}
	\right]
	=
	\int d^{p+1}\xi_{b}
	\left[
	(\frac{\partial}{\partial X_{a}^{\mu_0}}
	- \frac{1}{p!}D_a^{\mu_{1}\mu_{2}\ldots\mu_{p}}
	\frac{\partial}%
	{\partial \sigma_a^{\mu_{0}\mu_{1}\ldots\mu_{p}}})R_{ab}
	\right],                                                \label{eq:mo-2}
\end{equation}
and Eq.(\ref{eq:sur-1}) as\footnote{
Here care should be taken of the differentiation
with respect to the antisymmetric tensor elements
$\sigma^{\mu_{0}\mu_{1}\ldots\mu_{p}}$.
One is to differentiate in
$\sigma^{\mu_{0}\mu_{1}\ldots\mu_{p}}$,
with every permutation of $\mu_{1}\mu_{2}\ldots\mu_{p}$.}
\begin{equation}
	\frac{\sigma_{a\mu_{0}\mu_{1}\ldots\mu_{p}}}%
	{(-\sigma_{a}\cdot\sigma_{a})^{\frac{1}{2}}}
	K_{ai}^{\mu_{1}\mu_{2}\ldots\mu_{p}}
	=
	- \frac{1}{(p+1)!}
	\int d^{p+1}\xi_{b}K_{ai}^{\mu_{1}\mu_{2}\ldots\mu_{p}}
	\frac{\partial R_{ab}}%
	{\partial \sigma_a^{\mu_{0}\mu_{1}\ldots\mu_{p}}},     \label{eq:sur-2}
\end{equation}
\begin{displaymath}
	(\xi_a^i = 0,l_a^i).
\end{displaymath}

Assume a concrete form of $R_{ab}$
just in the same way as in the case of closed p-branes:
\begin{equation}
	R_{ab}
	\equiv
	g_{a}g_{b}
	\sigma_{a}^{\mu_{0}\mu_{1}\ldots\mu_{p}}
	\sigma_{b\mu_{0}\mu_{1}\ldots\mu_{p}}
	G((X_{a}-X_{b})^2),                                       \label{def:R}
\end{equation}
with $g_{a}$ and $g_{b}$, coupling constants,
and $G$ is some propagator in $D$-dimensional space-time.
Substitution of Eq.(\ref{def:R})
into Eqs.(\ref{eq:mo-2}) and (\ref{eq:sur-2})
leads us to
\begin{equation}
	(p+1)D_a^{\mu_{1}\mu_{2}\ldots\mu_{p}}
	\left[
	\frac{\sigma_{a\mu_{0}\mu_{1}\ldots\mu_{p}}}%
	{(-\sigma_{a}\cdot\sigma_{a})^{\frac{1}{2}}}
	\right]
	=
	g_{a}\sigma_{a}^{\mu_{1}\mu_{2}\ldots\mu_{p+1}}
	F_{b\mu_{0}\mu_{1}\ldots\mu_{p+1}}(X_a),                \label{eq:mo-3}
\end{equation}
and
\begin{equation}
	\frac{\sigma_{a\mu_{0}\mu_{1}\ldots\mu_{p}}}%
	{(-\sigma_{a}\cdot\sigma_{a})^{\frac{1}{2}}}
	K_{ai}^{\mu_{1}\mu_{2}\ldots\mu_{p}}
	=
	-g_{a}K_{ai}^{\mu_{1}\mu_{2}\ldots\mu_{p}}
	\phi_{b\mu_{0}\mu_{1}\ldots\mu_{p}}(X_a),              \label{eq:sur-3}
\end{equation}
\begin{displaymath}
	(\xi_{a}^{i}=0,l_{a}^{i}),
\end{displaymath}
where $F_{b}$ and $\phi_{b}$ are antisymmetric tensors defined by
\begin{eqnarray}
	F_b^{\mu_{0}\mu_{1}\ldots\mu_{p+1}}(X)
	& \equiv &
	\sum_{i=0}^{p+1}(-1)^{(p+1)\cdot i}
	\frac{\partial}{\partial X_{\mu_{i}}}
	\phi_{b}^{\mu_{i+1}\mu_{i+2}\ldots\mu_{p+1}
	\mu_{0}\mu_{1}\ldots\mu_{i-1}}
	(X),                                                      \label{def:F}
	\\
	\phi_{b}^{\mu_{0}\mu_{1}\ldots\mu_{p}}(X)
	& \equiv &
	g_{b}\int d^{p+1}\xi_{b}\sigma_{b}^{\mu_{0}\mu_{1}\ldots\mu_{p}}
	G((X-X_b)^2).                                           \label{def:phi}
\end{eqnarray}

Now, let us investigate the conservation of a current for field $\phi_{b}$
and the gauge invariance of the action.

First, we consider the conservation of a current.
In general,
$G$ obeys, being a propagator,
\begin{equation}
	\left(
	\frac{\partial^2}{\partial X_{\mu}\partial X^{\mu}}+m^2
	\right)
	G(X^2)
	=
	-C\delta^{(D)}(X),                                      \label{G:delta}
\end{equation}
where $C$ is a dimensionless constant
and $m$ is to be identified with the mass of the field.
By Eq.(\ref{G:delta}), we obtain
\begin{equation}
	\left(
	\frac{\partial^2}{\partial X_{\mu}\partial X^{\mu}}+m^2
	\right)
	\phi_{b}^{\mu_{0}\mu_{1}\ldots\mu_{p}}(X)
	=
	-Cj_{b}^{\mu_{0}\mu_{1}\ldots\mu_{p}}(X),               \label{phi:cur}
\end{equation}
with
\begin{equation}
	j_{b}^{\mu_{0}\mu_{1}\ldots\mu_{p}}(X)
	\equiv
	g_{b}\int d^{p+1}\xi_{b}\sigma_{b}^{\mu_{0}\mu_{1}\ldots\mu_{p}}
	\delta^{(D)}(X-X_{b}).                                  \label{def:cur}
\end{equation}
We further find,
with a bit of algebra, that
\begin{equation}
	\frac{\partial}{\partial X^{\mu_0}}
	\phi_{b}^{\mu_{0}\mu_{1}\ldots\mu_{p}}(X)
	=
	- g_{b}\sum_{i=0}^{p}
	\int d\xi^0_{b}d\xi^1_{b}\ldots\hat{d\xi^{i}_{b}}\ldots d\xi^{p}_{b}
	\left[
	K_{bi}^{\mu_{1}\mu_{2}\ldots\mu_{p}}G((X-X_{b})^2)
	\right]_{\xi_{b}^{i}=0}^{\xi_{b}^{i}=l_{b}^{i}}.        \label{div:phi}
\end{equation}
Here $\hat{d\xi^{i}}$ means exclusion of $d\xi^{i}$.
On account of the existence of boundary surface,
the divergence of $\phi_{b}$ does not vanish.
{}From Eqs.(\ref{phi:cur}) and (\ref{div:phi}) follows the equation
\begin{equation}
	\frac{\partial}{\partial X^{\mu_0}}
	j_{b}^{\mu_{0}\mu_{1}\ldots\mu_{p}}(X)
	=
	- g_{b}\sum_{i=0}^{p}
	\int d\xi^0_{b}d\xi^1_{b}\ldots\hat{d\xi^{i}_{b}}\ldots d\xi^{p}_{b}
	\left[
	K_{bi}^{\mu_{1}\mu_{2}\ldots\mu_{p}}
	\delta^{(D)}(X-X_b)
	\right]_{\xi_b^i=0}^{\xi_b^i=l_b^i},                    \label{div:cur}
\end{equation}
which shows that
the current $j_{b}$ would be conserved
if the p-brane was closed.
The modification of the current for the open p-brane
will be done after the introduction of a gauge transformation.

Secondly, we examine the gauge invariance of the action,
defining a gauge transformation by
\begin{equation}
	\phi^{\mu_{0}\mu_{1}\ldots\mu_{p}}(X)
	\rightarrow
	\phi^{\mu_{0}\mu_{1}\ldots\mu_{p}}(X)
	+ \sum_{i=0}^{p}(-1)^{p\cdot i}
	\frac{\partial}{\partial X_{\mu_{i}}}
	\Lambda^{\mu_{i+1}\mu_{i+2}\ldots\mu_{p}
	\mu_{0}\mu_{1}\ldots\mu_{i-1}},                          \label{gauge1}
\end{equation}
with an arbitrary antisymmetric tensor $\Lambda$.
Instead of the whole action $S$,
we take up
\begin{eqnarray}
	S^{\rm int}
	& \equiv &
	\int d^{p+1}\xi_{a}\int d^{p+1}\xi_{b}R_{ab}               \nonumber \\
	& = &
	g_{a}\int d^{p+1}\xi_{a}\sigma_{a}^{\mu_{0}\mu_{1}\ldots\mu_{p}}
	\phi_{b\mu_{0}\mu_{1}\ldots\mu_{p}},                   \label{def:Sint}
\end{eqnarray}
{\it i.e.},
the $\phi$-depending part $S^{\rm int}$ of $S$.
Under the transformation (\ref{gauge1}),
$S^{\rm int}$ transforms as follows:
\begin{eqnarray}
	S^{\rm int}
	& \rightarrow &
	S^{\rm int}
	+ (p+1)g_{a}\int d^{p+1}\xi_{a}D_{a}^{\mu_{1}\mu_{2}\ldots\mu_{p}}
	\Lambda_{\mu_{1}\mu_{2}\ldots\mu_{p}}             \label{Gvar:Sint1} \\
	& = &
	S^{\rm int}
	+ (p+1)g_{a}
	\int d\xi^0_{a}d\xi^1_{a}\ldots\hat{d\xi^{i}_{a}}\ldots d\xi^{p}_{a}
	\left[
	\sum_{i=0}^{p}K_{ai}^{\mu_{1}\mu_{2}\ldots\mu_{p}}
	\Lambda_{\mu_{1}\mu_{2}\ldots\mu_{p}}
	\right]_{\xi_{a}^{i}=0}^{\xi_{a}^{i}=l_{a}^{i}}.     \label{Gvar:Sint2}
\end{eqnarray}
The second term of the right-hand side in Eq.(\ref{Gvar:Sint2})
does not vanish,
here also,
due to the existence of boundary surface.
Note that it vanishes for {\it closed} p-branes.
With Eq.(\ref{Gvar:Sint2}),
one immediately sees that $S^{\rm int}$ is not invariant
under the transformation (\ref{gauge1}).
Although the equations of motion (\ref{eq:mo-3}) are invariant
under this gauge transformation,
the equations of boundary surface (\ref{eq:sur-3}) are not.
The latter become
\begin{equation}
	\frac{\sigma_{a\mu_{0}\mu_{1}\ldots\mu_{p}}}%
	{(-\sigma_{a}\cdot\sigma_{a})^{\frac{1}{2}}}
	K_{ai}^{\mu_{1}\mu_{2}\ldots\mu_{p}}
	=
	-g_{a}K_{ai}^{\mu_{1}\mu_{2}\ldots\mu_{p}}
	\left[
	\phi_{b\mu_{0}\mu_{1}\ldots\mu_{p}}(X_a)
	+ \sum_{i=0}^{p}(-1)^{p\cdot i}
	\frac{\partial}{\partial X^{\mu_{i}}}
	\Lambda_{\mu_{i+1}\mu_{i+2}\ldots\mu_{p}\mu_{0}\mu_{1}\ldots\mu_{i-1}}
	\right],                                            \label{Gvar:eq:sur}
\end{equation}
\begin{displaymath}
	(\xi_{a}^{i}=0,l_{a}^{i}).
\end{displaymath}

To recover the invariance of the action,
we therefore add another term to the action $S$.
Based on Eqs.(\ref{Gvar:Sint2}) and (\ref{Gvar:eq:sur}),
we take up as an additional action
\begin{equation}
	S^{\rm add}
	\equiv
	-(p+1)e_{a}e_{b}\int d^{p+1}\xi_{a}\int d^{p+1}\xi_{b}
	D_{a}^{\mu_{1}\mu_{2}\ldots\mu_{p}}
	D_{b\mu_{1}\mu_{2}\ldots\mu_{p}}G((X_{}a-X_{b})^2).    \label{def:Sadd}
\end{equation}
It is straightforward,
under the variation in $X_{a}^{\mu}$,
that $S^{\rm add}$ does not change the equations of motion.
Furthermore, Eq.(\ref{def:Sadd}) is rewritten as follows:
\begin{equation}
	S^{\rm add}
	=
	(p+1)e_{a}\int d^{p+1}\xi_{a}
	D_{a}^{\mu_{1}\mu_{2}\ldots\mu_{p}}
	B_{b\mu_{1}\mu_{2}\ldots\mu_{p}},                         \label{Sadd2}
\end{equation}
with
\begin{equation}
	B_{b}^{\mu_{1}\mu_{2}\ldots\mu_{p}}(X)
	\equiv
	- e_b\sum_{i=0}^{p}
	\int d\xi^0_{b}d\xi^1_{b}\ldots\hat{d\xi^{i}_{b}}\ldots d\xi^{p}_{b}
	\left[
	K_{bi}^{\mu_{1}\mu_{2}\ldots\mu_{p}}G((X-X_{b})^2)
	\right]_{\xi_{b}^{i}=0}^{\xi_{b}^{i}=l_{b}^{i}}.          \label{def:B}
\end{equation}
Equations (\ref{Sadd2}) and (\ref{def:B}) trivially teach us that
$S^{\rm add}$ vanishes when p-branes $a$ and $b$ are closed.

In view of Eqs.(\ref{Gvar:Sint1}) and (\ref{Sadd2}),
we define a new gauge transformation
\begin{eqnarray}
	\phi^{\mu_{0}\mu_{1}\ldots\mu_{p}}(X)
	& \rightarrow &
	\phi^{\mu_{0}\mu_{1}\ldots\mu_{p}}(X)
	+ \sum_{i=0}^{p}(-1)^{p\cdot i}
	\frac{\partial}{\partial X_{\mu_{i}}}
	\Lambda^{\mu_{i+1}\mu_{i+2}\ldots\mu_{p}
	\mu_{0}\mu_{1}\ldots\mu_{i-1}},                            \nonumber \\
	B^{\mu_{1}\mu_{2}\ldots\mu_{p}}(X)
	& \rightarrow &
	B^{\mu_{1}\mu_{2}\ldots\mu_{p}}
	- \frac{g}{e}\Lambda^{\mu_{1}\mu_{2}\ldots\mu_{p}}.      \label{gauge2}
\end{eqnarray}
It is easy to see that
these transformations make the action $S+S^{\rm add}$ invariant.

Under the variation in $X_{a}^{\mu}$,
the action $S+S^{\rm add}$ yields the following equations of boundary surface
\begin{equation}
	\frac{\sigma_{a\mu_{0}\mu_{1}\ldots\mu_{p}}}%
	{(-\sigma_{a}\cdot\sigma_{a})^{\frac{1}{2}}}
	K_{ai}^{\mu_{1}\mu_{2}\ldots\mu_{p}}
	=
	-\left[
	g_{a}\phi_{b\mu_{0}\mu_{1}\ldots\mu_{p}}(X_{a})
	+e_{a}\sum_{i=0}^{p}(-1)^{p\cdot i}
	\frac{\partial}{\partial X_{a}^{\mu_{i}}}
	B_{b\mu_{i+1}\mu_{i+2}\ldots\mu_{p}\mu_{0}\mu_{1}\ldots\mu_{i-1}}
	\right]
	K_{ai}^{\mu_{1}\mu_{2}\ldots\mu_{p}},                  \label{eq:sur-4}
\end{equation}
\begin{displaymath}
	(\xi_{a}^{i}=0,l_{a}^{i}).
\end{displaymath}
These equations of boundary surface are invariant
under the transformations (\ref{gauge2}).

The Lagrangian density for the fields $\phi$ and $B$
is given by
\begin{eqnarray}
	{\cal L}
	& = &
	- \frac{1}{2(p+2)!}
	F^{\mu_{0}\mu_{1}\ldots\mu_{p+1}}
	F_{\mu_{0}\mu_{1}\ldots\mu_{p+1}}                          \nonumber \\
	& &
	+ \frac{1}{2(p+1)!}
	\left[
	\sum_{i=0}^{p}(-1)^{p\cdot i}
	\frac{\partial}{\partial X^{\mu_{i}}}
	B_{\mu_{i+1}\mu_{i+2}\ldots\mu_{p}\mu_{0}\mu_{1}\ldots\mu_{i-1}}
	\right]^2 \nonumber \\
	& &
	+ \frac{1}{2(p+1)!}\frac{g^2}{e^2}
	\phi^{\mu_{0}\mu_{1}\ldots\mu_{p}}
	\phi_{\mu_{0}\mu_{1}\ldots\mu_{p}}                         \nonumber \\
	& &
	+ \frac{1}{(p+1)!}\frac{g}{e}
	\phi^{\mu_{0}\mu_{1}\ldots\mu_{p}}
	\left[
	\sum_{i=0}^{p}(-1)^{p\cdot i}
	\frac{\partial}{\partial X^{\mu_{i}}}
	B_{\mu_{i+1}\mu_{i+2}\ldots\mu_{p}\mu_{0}\mu_{1}\ldots\mu_{i-1}}
	\right],                                            \label{lagrangian1}
\end{eqnarray}
which is also invariant under the transformations (\ref{gauge2}).
The variational principle for this Lagrangian ${\cal L}$
with respect to $\phi$ and $B$
leads to the equation of motion for $B$:
\begin{equation}
        \frac{\partial^2}{\partial X_{\mu_0}\partial X^{\mu_0}}
	B^{\mu_{1}\mu_{2}\ldots\mu_{p}}
	=
	- \sum_{i=1}^{p}(-1)^{p\cdot i}
	\frac{\partial}{\partial X_{\mu_{i}}}
	\frac{\partial}{\partial X^{\mu_0}}
	B^{\mu_{i+1}\mu_{i+2}\ldots\mu_{p}\mu_{0}\mu_{1}\ldots\mu_{i-1}}
        - \frac{g}{e}\frac{\partial}{\partial X^{\mu_0}}
	\phi^{\mu_{0}\mu_{1}\ldots\mu_{p}},                    \label{eq:mo-B1}
\end{equation}
and that for $\phi$:
\begin{eqnarray}
	\frac{\partial^2}{\partial X_{\mu_0}\partial X^{\mu_0}}
	\phi^{\mu_{1}\mu_{2}\ldots\mu_{p+1}}
	& = &
	- \sum_{i=1}^{p+1}(-1)^{(p+1)\cdot i}
	\frac{\partial}{\partial X_{\mu_{i}}}
	\frac{\partial}{\partial X^{\mu_0}}
	\phi^{\mu_{i+1}\mu_{i+2}\ldots\mu_{p+1}
	\mu_{0}\mu_{1}\ldots\mu_{i-1}}                             \nonumber \\
	& &
	- \frac{g}{e}\sum_{i=1}^{p+1}(-1)^{p\cdot (i+1)}
	\frac{\partial}{\partial X_{\mu_{i}}}
	B^{\mu_{i+1}\mu_{i+2}\ldots\mu_{p+1}
	\mu_{1}\mu_{2}\ldots\mu_{i-1}}                             \nonumber \\
	& &
	- \frac{g^2}{e^2}
	\phi^{\mu_{1}\mu_{2}\ldots\mu_{p+1}}.                \label{eq:mo-phi1}
\end{eqnarray}

We next investigate the conservation of a current for $B_{b}$.
Comparing Eq.(\ref{def:B}) with Eq.(\ref{div:phi}),
we immediately find that
$\phi_{b}$ and $B_{b}$ satisfy a gauge condition
\begin{equation}
	\frac{\partial}{\partial X^{\mu_0}}
	\phi_{b}^{\mu_{0}\mu_{1}\ldots\mu_{p}}(X)
	=
	\frac{g_b}{e_b}
	B_{b}^{\mu_{1}\mu_{2}\ldots\mu_{p}}(X).                   \label{phi:B}
\end{equation}
Furthermore, $B_{b}$ obeys
\begin{equation}
	\frac{\partial}{\partial X^{\mu_1}}
	B_{b}^{\mu_{1}\mu_{2}\ldots\mu_{p}}(X)
	=
	0.                                                        \label{div:B}
\end{equation}
Therefore, we define a new current $J_{b}$ for $B_{b}$ by
\begin{equation}
	J_{b}^{\mu_{1}\mu_{2}\ldots\mu_{p}}
	\equiv
	e_{b}\sum_{i=1}^{p}
	\int d\xi^0_{b}d\xi^1_{b}\ldots\hat{d\xi^{i}_{b}}\ldots d\xi^{p}_{b}
	\left[
	K_{bi}^{\mu_{1}\mu_{2}\ldots\mu_{p}}\delta^{(D)}(X-X_b)
	\right]_{\xi_{b}^{i}=0}^{\xi_{b}^{i}=l_{b}^{i}},        \label{def:Cur}
\end{equation}
by which we have
\begin{equation}
	\left(
	\frac{\partial^2}{\partial X_{\mu}\partial X^{\mu}}+m^2
	\right)
	B_{b}^{\mu_{1}\mu_{2}\ldots\mu_{p}}(X)
	=
	-CJ_{b}^{\mu_{1}\mu_{2}\ldots\mu_{p}}(X).                 \label{B:Cur}
\end{equation}
According to the vanishing of the divergence of $B_{b}$,
$J_{b}$ satisfies the conservation law of a current
\begin{equation}
	\frac{\partial}{\partial X^{\mu_1}}
	J_{b}^{\mu_{1}\mu_{2}\ldots\mu_{p}} = 0.                \label{div:Cur}
\end{equation}
It follows from
Eqs.(\ref{phi:cur}), (\ref{B:Cur}) and (\ref{phi:B}) that
\begin{equation}
	J_{b}^{\mu_{1}\mu_{2}\ldots\mu_{p}}(X)
	=
	\frac{e_{b}}{g_{b}}\frac{\partial}{\partial X^{\mu_0}}
	j_{b}^{\mu_{0}\mu_{1}\ldots\mu_{p}}(X).                 \label{Cur:cur}
\end{equation}
This shows that
the new conserved current $J_{b}$ instead of the nonconserved one $j_{b}$
is {\it physical}.
Note that the divergence of $j_{b}$ enters into Eq.(\ref{Cur:cur}).

With gauge conditions (\ref{phi:B}) and (\ref{div:B}),
Eq.(\ref{eq:mo-B1}) becomes
\begin{equation}
        \frac{\partial^2}{\partial X_{\mu_0}\partial X^{\mu_0}}
	B^{\mu_{1}\mu_{2}\ldots\mu_{p}}
        =
        -\left(\frac{g}{e}\right)^2
        B^{\mu_{1}\mu_{2}\ldots\mu_{p}},                       \label{eq:mo-B2}
\end{equation}
which means
$B$ is a field with mass
\begin{equation}
        m = \frac{g}{e}.                                           \label{mass}
\end{equation}
Similarly,
the equation of motion for $\phi$ is
\begin{equation}
	\frac{\partial^2}{\partial X_{\mu_0}\partial X^{\mu_0}}
	\phi^{\mu_{1}\mu_{2}\ldots\mu_{p+1}}
	=
	-\left(\frac{g}{e}\right)^2
	\phi^{\mu_{1}\mu_{2}\ldots\mu_{p+1}},                \label{eq:mo-phi2}
\end{equation}
and $\phi$ is considered as a field with mass $m=g/e$.

Now we combine $\phi$ with $B$
to introduce a new field
\begin{equation}
	\psi^{\mu_{0}\mu_{1}\ldots\mu_{p}}(X)
	\equiv
	\phi^{\mu_{0}\mu_{1}\ldots\mu_{p}}(X)
	+
	\frac{e}{g}\sum_{i=0}^{p}(-1)^{p\cdot i}
	\frac{\partial}{\partial X_{\mu_{i}}}
	B^{\mu_{i+1}\mu_{i+2}\ldots\mu_{p}
	\mu_{0}\mu_{1}\ldots\mu_{i-1}}(X).                      \label{def:psi}
\end{equation}
It is clear that
this field $\psi$ is invariant
under the transformations (\ref{gauge2}).
In terms of $\psi$,
we can rewrite ${\cal L}$ as
\begin{equation}
	{\cal L}
	=
	- \frac{1}{2(p+2)!}
	F^{\mu_{0}\mu_{1}\ldots\mu_{p+1}}
	F_{\mu_{0}\mu_{1}\ldots\mu_{p+1}}
	+ \frac{1}{2(p+1)!}\frac{g^2}{e^2}
	\psi^{\mu_{0}\mu_{1}\ldots\mu_{p}}
	\psi_{\mu_{0}\mu_{1}\ldots\mu_{p}}.                 \label{lagrangian2}
\end{equation}
The gauge invariance of $\psi$ also leads to the invariance of ${\cal L}$.

The gauge conditions (\ref{div:phi}) and (\ref{div:B})
induce, for $\psi$,
\begin{equation}
	\frac{\partial}{\partial X^{\mu_0}}
	\psi^{\mu_{0}\mu_{1}\ldots\mu_{p}} = 0.                 \label{div:psi}
\end{equation}
And the equation of motion for $\psi$ is obtained as
\begin{equation}
	\frac{\partial^2}{\partial X_{\mu_0}\partial X^{\mu_0}}
	\psi^{\mu_{1}\mu_{2}\ldots\mu_{p+1}}
	=
	-\left(\frac{g}{e}\right)^2
	\psi^{\mu_{1}\mu_{2}\ldots\mu_{p+1}}.                \label{eq:mo-psi2}
\end{equation}
It shows that $\psi$ is again a field with mass $m=g/e$.

Thus we have obtained a massive gauge-invariant field $\psi$.
The total action,
which describes the open p-brane and its interaction with the field $\psi$,
is given by
\begin{equation}
        S^{\rm total} =
        -
        \int d^{p+1}\xi (-\sigma\cdot\sigma)^{\frac{1}{2}}
        +
        g\int d^{p+1}\xi \sigma^{\mu_{0}\mu_{1}\ldots\mu_{p}}
	\psi_{\mu_{0}\mu_{1}\ldots\mu_{p}}
        +
        \int d^{D}X {\cal L}.                                    \label{Action}
\end{equation}

\vspace*{8.0mm}

Now we come to our conclusions.
As we have discussed above,
it is the existence of the boundary surface that
makes a difference between closed and open p-branes.
Closed p-branes are described by the equations of motion (\ref{eq:mo-3})
without equations of boundary surface.
Consequently, their interactions are regarded as
mediated by the massless field $\phi$
defined by Eq.(\ref{def:phi})\cite{yama}.
But when $\phi$ is introduced in case of open p-branes,
two difficulties occur
owing to this existence of the boundary surface of world volumes.
The one is that the current $j$, defined by Eq.(\ref{def:cur}),
for $\phi$ is not to be conserved,
the other is that the action $S$ is not gauge invariant
under the transformation (\ref{gauge1}).
To overcome these difficulties
we have added the new term $S^{\rm add}$ to the action $S$,
and defined the new gauge transformations by Eq.(\ref{gauge2}).
The introduction of them brings us to that of a new field $B$.
This further enforces us to have the action $S+S^{\rm add}$ instead of $S$,
by which the action has been recovered its gauge invariance.
We find that the current $J$ for the field $B$,
which is regarded as the divergence of the current $j$,
is conserved with Eqs.(\ref{div:Cur}) and (\ref{Cur:cur}).

As a consequence,
we have obtained the field $\psi$,
which is constructed by the combination of the fields $\phi$ and $B$,
as a medium of interaction of the open p-branes.
In contrast to the case of the closed p-branes
the field $\psi$ has clearly a mass $m=g/e$;
besides,
it is gauge invariant under the transformation (\ref{gauge2}).

\vspace{8.0mm}

One of the authors (M.Y.) would like to
thank Iwanami F\^ujukai for financial support.

\end{document}